\documentclass[12pt]{article}
\usepackage{scicite}
\usepackage{times}
\usepackage{graphicx}

\newenvironment{sciabstract}{%
\begin{quote} \bf}
{\end{quote}}

\newcounter{lastnote}

\title{Quantum-state resolved bimolecular collisions of velocity-controlled OH with NO radicals}

\author
{Moritz Kirste,$^{1\ast}$ Xingan Wang,$^{1\ast}$ H. Christian Schewe,$^1$ Gerard Meijer,$^{1}$ Kopin Liu,$^2$ \\Ad van der Avoird,$^3$ Liesbeth M.C. Janssen,$^3$ Koos B. Gubbels,$^3$ \\ Gerrit C. Groenenboom,$^{3\ast\ast}$ Sebastiaan Y.T. van de Meerakker,$^{3,1\ast\ast}$\\  \\
\\
\normalsize{$^{1}$Fritz-Haber-Institut der
Max-Planck-Gesellschaft, Faradayweg 4-6,
14195 Berlin, Germany}\\
\normalsize{$^{2}$Institute of Atomic and Molecular Sciences (IAMS), Academia Sinica, Taipei, Taiwan 10617}\\
\normalsize{$^{3}$Radboud University Nijmegen, Institute for Molecules and Materials}\\
\normalsize{Heijendaalseweg 135, 6525 AJ Nijmegen, the Netherlands}\\
\\
\normalsize{$^{\ast}$Who contributed equally to this work;}\\
\normalsize{$^{\ast\ast}$To whom correspondence should be addressed;}\\
\normalsize{E-mail: basvdm@science.ru.nl,
gerritg@theochem.ru.nl} }

\date{}

\begin{document}

\baselineskip24pt

\maketitle

\begin{sciabstract}
Whereas atom-molecule collisions have been studied with complete quantum state resolution, 
interactions between two state-selected molecules have proven much harder to probe. Here, we report the
measurement of state-resolved inelastic scattering cross sections for
collisions between two open-shell molecules that are both prepared in a
single quantum state. Stark-decelerated OH radicals were scattered with
hexapole-focused NO radicals in a crossed beam configuration.
Rotationally and spin-orbit inelastic scattering cross sections were
measured on an absolute scale for collision energies between 70 and 300
cm$^{-1}$. These cross sections show fair agreement with quantum
coupled-channels calculations using a set of coupled model potential
energy surfaces based on \textit{ab initio} calculations for the
long-range non-adiabatic interactions and a simplistic
short-range interaction. This comparison reveals the crucial role of electrostatic
forces in complex molecular collision processes.
\end{sciabstract}%

Rotationally inelastic scattering is one of the key processes underlying the exchange of energy between molecules \cite{Levine:reaction-dynamics,Chandler:book}. In bulk systems, rotational energy transfer (RET) is responsible for the thermalization of state populations following a chemical reaction. In the dilute interstellar medium, inelastic collisions contribute to the formation of non-thermal population distributions that result in, for instance, interstellar masers \cite{Weinreb:Nature200:829}. Accurate state-to-state inelastic scattering cross sections are essential ingredients for reliable models of chemical processes in combustion physics, atmospheric science, and astrochemistry.

In molecular beam collision experiments, the ability to prepare molecules in a single rotational (sub)level prior to the collision using electric, magnetic, or optical fields has been imperative to unravel the underlying mechanisms of molecular energy transfer. This has made scattering experiments possible at the full state-to-state level, and has resulted in the discovery of propensity rules for inelastic scattering \cite{Schiffman:IRPC14:1995}, the stereodynamics of molecular collisions \cite{Stolte:NAT353:391,Watanabe:PRL99:043201}, and quantum interference effects \cite{Lorenz:SCIENCE293:2063,Eyles:NatChem3:597,Kohguchi:Science294:832}. The latest beam deceleration and acceleration methods \cite{Meerakker:NatPhys4:595,Meerakker:ChemRev:inpress} allow for the precise variation of the collision energy, resulting in the observation of quantum threshold effects in the state-to-state cross sections \cite{Gilijamse:Science313:1617,Scharfenberg:PCCP12:10660}. This wealth of studies has contributed enormously to our present understanding of how intermolecular potentials govern molecular collision dynamics.

Thus far these methods have mostly been used to study collisions of state-selected molecules with rare gas atoms. Yet, in most natural environments molecule-molecule interactions play a major role. For instance, space telescope observations of cometary water may reveal the possible origin of water on Earth, but a conclusive interpretation requires accurate knowledge of RET in water-water collisions \cite{Hartogh:Nature478:218}. Whereas atom-molecule scattering cross sections can now be calculated routinely in excellent agreement with experiment \cite{Scharfenberg:PCCP12:10660,Paterson:IRPC31:69}, much less is known about RET in molecule-molecule collisions \cite{Clary:ARPC41:61}. As opposed to an atomic target, a molecular scattering partner possesses internal degrees of freedom of its own, adding a level of complexity that can easily render \emph{ab initio} quantum scattering calculations extremely challenging -- if not impossible. This is particularly true for collisions involving radical species that are governed by multiple Born-Oppenheimer (BO) potential energy surfaces (PESs) with non-adiabatic couplings between them. Experimental data on bimolecular state-to-state cross sections is generally lacking, and kinetic models often use collision rate coefficients that are expected to be inaccurate \cite{Smith:ARAA49:29}.

The study of molecule-molecule collisions at the ultimate quantum level has been a quest in molecular beam physics since it was established in the 1950s \cite{Bernstein:Science144:141}. Major obstacles exist that have prevented studies of state-to-state bimolecular scattering \cite{Sawyer:PCCP13:19059}. The main challenge is the need for reagent beams with sufficient quantum state purity at the densities necessary to observe population transfer in one, or both, reagent beam(s). Thus far, experiments of this kind have only been possible using cryogenically cooled H$_2$ molecules as a target beam \cite{Berteloite:PRL105:203201,Schreel:JCP105:4522}.

Here, we report the successful measurement of state-resolved inelastic scattering
between two state-selected molecular beams. We have chosen the OH
($X\,^2\Pi$) + NO ($X\,^2\Pi$) system \cite{Note:1} for this purpose, as
both open shell radical species are benchmark systems for the scattering
of state-selected molecules with rare gas atoms that involve two BO PESs
\cite{Kohguchi:ARPC98:421}. Collisions between OH and NO involve
eight interacting PESs, representing the full complexity of bimolecular
inelastic collisions \cite{Vonk:JCP106:1353}. The OH-NO system serves
also as a prototypical example of radical-radical reactions
of fundamental importance in gas-phase chemical kinetics
\cite{Sharkey:JCSFT90:3609}. We used a Stark-decelerator and a hexapole
state selector in a crossed molecular beam configuration to produce
reagent beams of OH and NO radicals with an almost perfect quantum state
purity. The collision energy was varied between 70 and 300 cm$^{-1}$ by
tuning the velocity of the OH radicals prior to the collision using the
Stark decelerator, revealing the quantum threshold behavior of the
state-to-state inelastic scattering cross sections. The unusually well
defined distributions of reagent molecules allowed us to determine
absolute scattering cross sections, which can normally be determined
only on a relative scale in crossed beam experiments. These cross
sections showed fair agreement with a theoretical model for inelastic
collisions between two $^2\Pi$ radical species, based solely on an
accurate description of the full rotational and open-shell structure of
both radical species and their long-range non-adiabatic electrostatic
interactions. This study revealed that inelastic scattering
predominantly occurs at large intermolecular distances, even for the
relatively high collision energies probed here.

The crossed molecular beam apparatus used to study inelastic collisions between OH and NO radicals is schematically shown in Fig. 1 \cite{Note:SI}. A packet of OH radicals [$X\,^2\Pi_{3/2}, v=0, j=3/2, f$ \cite{Note:1}, referred to hereafter as $F_1(3/2f)$] with a tunable velocity in the 200 to 750 m/s range was produced using a 2.6-meter long Stark decelerator \cite{Scharfenberg:PRA79:023410}. The velocity of the OH radicals was tuned by applying a burst of high voltage pulses to the electric field electrodes at the appropriate times. The state purity of the OH packets was such that less than 0.01 \% of the OH radicals populated a lower $\Lambda$-doublet component of any rotational level.

A beam of NO radicals with a fixed velocity was produced by seeding NO in a xenon carrier gas, and was passed through a 30~cm long electrostatic hexapole. NO radicals in the low-field-seeking $F_1(1/2f)$ state were focused into the collision region, while molecules in the high-field-seeking $F_1(1/2e)$ state were deflected from the beam axis. A 2~mm diameter beamstop and diaphragm were installed in the center of the hexapole and 10~mm downstream from the hexapole, respectively, effectively filtering out the Xe atoms from the molecular beam pulse. The resulting state purity of the transmitted NO $F_1(1/2f)$ beam was better than 99.0 \%.

The reagent beams of OH and NO were detected state-selectively in the collision region by a laser induced fluorescence (LIF) detection scheme. From calibrated LIF measurements, the peak densities of the reagent packets were determined to be $(2 \pm 0.8) \times$ 10$^{8}$ and $(9 \pm 3) \times$ 10$^{10}$ molecules/cm$^{3}$ for OH and NO, respectively. The collision induced populations in the $F_1(3/2e)$, $F_1(5/2e)$, $F_1(7/2e)$, and the $F_2(1/2e)$ levels of the OH radical were measured at the time when both beams maximally overlapped in the beam crossing area. Depending on the inelastic channel, only a fraction of 10$^{-4}$ to 10$^{-6}$ of the OH radicals were inelastically scattered. Only final states of $e$ symmetry were probed, as the Stark decelerator did not eliminate the initial population in the $f$ states sufficiently. Detection of collision induced population in the $F_1(3/2e)$ level was only possible by spectroscopically separating the magnetic dipole allowed transitions that originated from the $F_1(3/2f)$ state \cite{Note:SI}. The insufficiently perfect state purity of the NO radical beam prevented the measurement of population transfer in NO.

The collision signals were measured as a function of the collision energy, from which the excitation functions of the state-to-state inelastic scattering cross sections (shown in Fig.\ 2) were determined \cite{Note:4}. The extremely well defined spatial distributions of the OH and NO packets allowed us to determine the scattering cross sections on an absolute scale. A cross section of 90~$\pm$ 38 $\textrm{\AA}^2$ was determined for the $F_1(3/2e)$ channel at a collision energy of 220~cm$^{-1}$, from which the absolute cross sections for all scattering channels and all collision energies were derived \cite{Note:SI}. The experimental uncertainty was limited only by the uncertainty in the measured value for the peak density of the NO packet.

We found that collisions that populate the $F_1(3/2e)$ level are most likely, and the cross section for this transition accounts for about 90 \% of the total inelastic scattering cross section. The cross sections to populate the $F_1(5/2e)$, the $F_1(7/2e)$, and the $F_2(1/2e)$ levels show a clear threshold behavior; the collision energies at which these channels become energetically possible are indicated by vertical arrows. These cross sections show large qualitative differences compared to the scattering of OH with atomic targets. The most striking difference is found in the relative contributions of the $F_1(3/2e)$ and $F_1(5/2e)$ channels to the total inelastic scattering cross section. The role of the $F_1(5/2e)$ channel, which dominates RET for OH-He and OH-Ne, gradually diminishes in favor of the $F_1(3/2e)$ channel in the series of targets He, Ne, Ar, Kr, and Xe \cite{Scharfenberg:EPJD65:189}. This behavior could be rationalized from the increasing well depth, anisotropy, and head-tail asymmetry of the two BO PESs \cite{Dagdigian:JCP130:094303,Scharfenberg:EPJD65:189}. The overwhelming dominance of the $F_1(3/2f) \rightarrow F_1(3/2e)$ quenching channel observed here for OH-NO reflects how a dipolar open-shell molecular scattering partner rather than a spherical atomic partner governs the collision dynamics.

In order to interpret our experimental results, we constructed a model
for the scattering of two molecules in an open shell $^2\Pi$ state. In
contrast to scattering of OH or NO with rare gas atoms, \emph{ab initio}
calculations of multiple anisotropic PESs with their non-adiabatic
couplings for OH-NO are beyond the capabilities of current theoretical
methods. Coupling of the $S=1/2$ electron spins gives rise to singlet
($S=0$) and triplet ($S=1$) potentials, which describe different
short-range exchange interactions. There are four spatially distinct
electronic states for each spin state, which are degenerate at long
range and for linear geometries and which are coupled by non-adiabatic
interactions. Nuclear derivative couplings with respect to all nuclear
degrees of freedom exist between these states. \textit{Ab initio}
studies of the OH-NO complex \cite{Nguyen:CP230:1} focused on the region
where the chemical reaction OH + NO $\rightarrow$ H + NO$_2$ takes
place, but considered only the lowest adiabatic potential for the
singlet state. Even if we were able to compute all the relevant
adiabatic PESs, there would be no simple recipe to take the
non-adiabatic couplings between the PESs into account.

In our model we exploit the hypothesis that the processes with the
largest cross sections are governed by couplings that occur at
relatively large OH-NO separations, beyond the HONO well region. As
opposed to the short range interactions, the long-range parts of the
PESs can be calculated accurately by \emph{ab initio} methods. We
neglected the complicated short-range behavior of the PESs and replaced it
with an isotropic repulsion term. However, we accurately calculated the
long-range PESs that are governed by first-order electrostatic
interactions between the dipole, quadrupole, and octupole moments of the
collision partners \cite{Note:SI}. Moreover, we included isotropic
dispersion and induction terms. The intermolecular Hamiltonian contained
the usual radial and centrifugal kinetic energy operators, and the full
$4 \times 4$ matrix of diabatic interaction potentials
\cite{Note:SI,Note:7}. Due to the non-cylindrical symmetry of the
$^2\Pi$ ground states of both the OH and NO radical, the off-diagonal
elements of this matrix provided by the quadrupole and octupole moments
of both radicals contain important couplings between the $^2\Pi_{3/2}$
and $^2\Pi_{1/2}$ states of both species.

The cross sections that were obtained from these model PESs by full
coupled-channels calculations are shown as solid lines in Figure 2. Fair agreement between experiment and theory
was obtained, in particular considering the simplistic approximations
for the short-range PESs that were made. The absolute value for the
cross section of the dominating $F_1(3/2e)$ channel, as well as the
relative strengths of the inelastic channels, are reproduced well by the
model calculations. The cross section for the spin-orbit changing
$F_2(1/2e)$ channel, as well as the cross section for the $F_1(3/2e)$
channel at low collision energies, are overestimated by the model.

We tested the sensitivity of the model calculations with respect to
changes in the short-range repulsion term \cite{Note:SI}. We observed
that the cross section for the parity-changing $F_1(3/2e)$ channel is
governed exclusively by the long-range electrostatic interaction; its
value is converged to within a few percent. About half of this large
quenching cross section originates from collisions with impact
parameters exceeding 12 $a_0$. The $F_1(5/2e)$ channel is also
mainly determined by the long-range forces, although its cross section
varies by 10 to 25 \% upon changes in the short-range model parameters
\cite{Note:SI}. The weak $F_1(7/2e)$ and $F_2(1/2e)$ channels show
larger variations, and more realistic short-range PESs are required to
accurately predict their cross sections.

Our model also predicts the final states of the NO radical that are
populated in coincidence with RET in the OH radical, but that cannot be
probed with the present experimental arrangement. The dominant
$F_1(3/2e)$ quenching transition in OH is accompanied by the inelastic
channels in NO as given in Table 1 for various collision energies. The
general scattering behavior can be understood from the terms that lead
to inversion parity changing or conserving collisions with respect to
both collision partners. The OH-NO dipole-dipole interaction results in
transitions that either change or conserve inversion parity in both OH
and NO, whereas the dipole-quadrupole and quadrupole-dipole terms also
allow for an inversion parity changing transition in OH or NO only.
The largest cross sections that accompany the $F_1(3/2e)$
channel in OH are found for the dipole-dipole dominated $F_1(3/2f)$ and
$F_1(1/2e)$ channels of NO, and for the dipole-quadrupole dominated
$F_1(1/2f)$, $F_1(3/2e)$, and $F_1(5/2f)$ NO channels \cite{Note:8}.

\begin{table}
\caption{Predicted state-to-state cross sections (\AA$^2$) for RET in NO
(initial state $F_1(1/2f)$) that occurs in coincidence with the
$F_1(3/2f) \rightarrow F_1(3/2e)$ transition in the OH radical at
collision energies of 10, 20, 50, 100, 200, and 300 cm$^{-1}$. The
inversion parity of the molecular levels is given. For comparison, the elastic 
cross section corresponding to the OH ($F_1(3/2f) \rightarrow F_1(3/2f)$) -- NO ($F_1(1/2f)\rightarrow F_1(1/2f)$) 
channel is given in the last row of the table.}
\begin{center}
\begin{tabular}{cccccccc}
final state & inversion & 10 & 20 & 50 & 100 & 200 & 300  \\
   (NO)     &   parity  & cm$^{-1}$ & cm$^{-1}$ & cm$^{-1}$ & cm$^{-1}$ & cm$^{-1}$ & cm$^{-1}$ \\   \hline
  $F_1(1/2e)$ & + & 152.3 & 106.6 & 54.5 & 33.5 & 20.0 & 15.3 \\
  $F_1(1/2f)$ & - & 45.1  & 37.0  & 25.8 & 13.9 & 6.3  & 4.2  \\
  $F_1(3/2e)$ & - & 19.1  & 17.6  & 21.4 & 17.5 & 12.1 & 10.0 \\
  $F_1(3/2f)$ & + & 18.0  & 19.5  & 46.9 & 54.3 & 38.8 & 29.7 \\
  $F_1(5/2e)$ & + &       & 6.5   & 8.5  & 9.0  & 6.3  & 5.0  \\
  $F_1(5/2f)$ & - &       & 6.6   & 7.1  & 8.7  & 11.8 & 13.4 \\
  $F_1(7/2e)$ & - &       &       & 5.3  & 5.4  & 4.4  & 3.8  \\
  $F_1(7/2f)$ & + &       &       & 4.7  & 5.2  & 5.6  & 5.6  \\
  $F_2(3/2e)$ & - &       &       &      &      & 0.5  & 0.4  \\
  $F_2(3/2f)$ & + &       &       &      &      & 0.5  & 0.4  \\
  $F_2(5/2e)$ & + &       &       &      &      & 0.4  & 0.3  \\
  $F_2(5/2f)$ & - &       &       &      &      & 0.4  & 0.4  \\
  $F_2(7/2e)$ & - &       &       &      &      & 0.3  & 0.3  \\
  $F_2(7/2f)$ & + &       &       &      &      & 0.3  & 0.3  \\ \hline
  elastic     &   & 409.6 & 346.3 & 265.1& 209.1& 175.8& 166.4 \\
\end{tabular}
\label{tab:pars}
\end{center}
\end{table}

Our experiments show that the
main mechanisms of RET in the reactive, chemically complex system studied here are
captured using a model for the long-range interactions alone, provided
that the full monomer Hamiltonians and all relevant long-range
non-adiabatic couplings are taken into account. Even at relatively high
collision energies, the inelastic scattering events with the largest
cross sections predominantly occur at large intermolecular distances where
the interaction potentials can be calculated accurately. The success
attained here implies that reliable predictions for state-to-state
scattering cross sections can now be made more generally for complex molecular systems
involving radicals, helping to solve urgent scientific questions in, for
instance, astrochemistry. Ultimately, new electronic structure methods
that include the chemically reactive short-range potentials and non-adiabatic couplings are required
to elucidate the exact mechanisms of radical-radical collisions. 


\begin{thebibliography}{10}
\bibitem{Note:1}
The labels $X^2\Pi_{3/2}, X^2\Pi_{1/2}, v$, and $j$ indicate the electronic
  states, the vibrational state and rotational state of both the OH and the NO
  radical, respectively. The spectroscopic symmetry labels $e$ and $f$ refer to
  the total parity of the electronic wavefunction, exclusive of rotation. The
  total inversion parity is indicated by the additional labels $+$ and $-$.
\bibitem{Note:7}
The OH-NO spin states with $S=0$ and $S=1$ are not distinguished in our model;
  the $4 \times 4$ matrices for each spin state are therefore identical.
\bibitem{Note:8}
Although the quadrupole moment vanishes in the $j=1/2$ state of NO, the
  quadrupole moment in the molecular frame contributes to the collisions
  producing NO states with $j=3/2$ and 5/2 that yield important contributions
  to the measured cross sections.
\end{thebibliography}

\bibliographystyle{Science}

\noindent S.Y.T.v.d.M. acknowledges support from the Netherlands Organisation for Scientific Research (NWO) via a VIDI grant. G.M. and K.B.G. acknowledge support from the ERC-2009-AdG under grant agreement 247142-MolChip. K.L. and A.v.d.A. acknowledge the Alexander von Humboldt Foundation (AvHF) for a Humboldt Research Award. X.W. acknowledges the AvHF for a research fellowship. We thank Janneke Blokland for her help setting up the narrowband laser system. We thank the referees for valuable and stimulating comments. The authors declare no competing financial interests.\\ 

\noindent
\textbf{Supplementary Materials}\\
Supplementary text\\
Figs. S1 to S3\\
Tables S1 and S2\\
References (34-49)\\

\begin{figure}
 \centering
 \includegraphics[width=0.9\columnwidth]{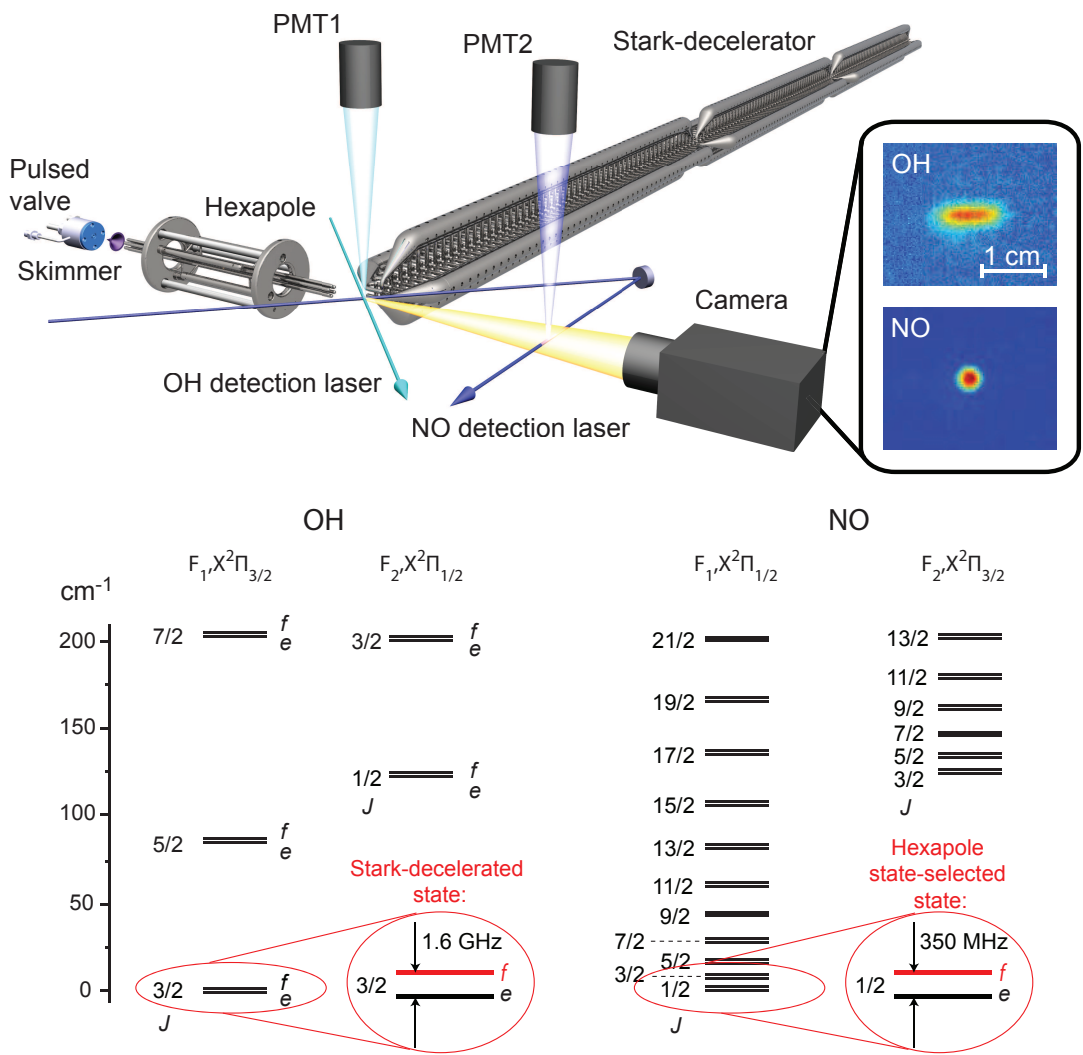}
\caption{Schematic representation of the experimental setup and the energy level schemes of the OH and NO radicals. A state-selected and velocity tunable beam of OH radicals produced using a 2.6-meter-long Stark decelerator was crossed with a hexapole state-selected beam of NO radicals. Both radical species were detected state-selectively using laser induced fluorescence (LIF), with total fluorescence intensity measured using a photomultiplier tube (PMT), and the spatial distribution of both reagent molecular packets imaged onto a charge-coupled device camera. Typical images of the OH and NO packets are shown in the upper and lower insets, respectively. The mean speed of the OH radical packet was precisely known from the settings of the Stark decelerator. The collision energy was calibrated from the NO beam speed measured via a second LIF detection zone located 30~cm downstream from the collision area. The $X\,^2\Pi$ electronic ground states of the OH and NO radicals are split into two rotational manifolds due to the spin-orbit interaction. The manifolds with lowest energy [$|\Omega|=3/2$ for OH and $|\Omega|=1/2$ for NO] are labeled $F_1$. The energy splittings shown between the $\Lambda$-doublet components of each rotational level are greatly exaggerated for clarity.}
\end{figure}

\begin{figure}
 \centering
 \includegraphics[width=0.9\columnwidth]{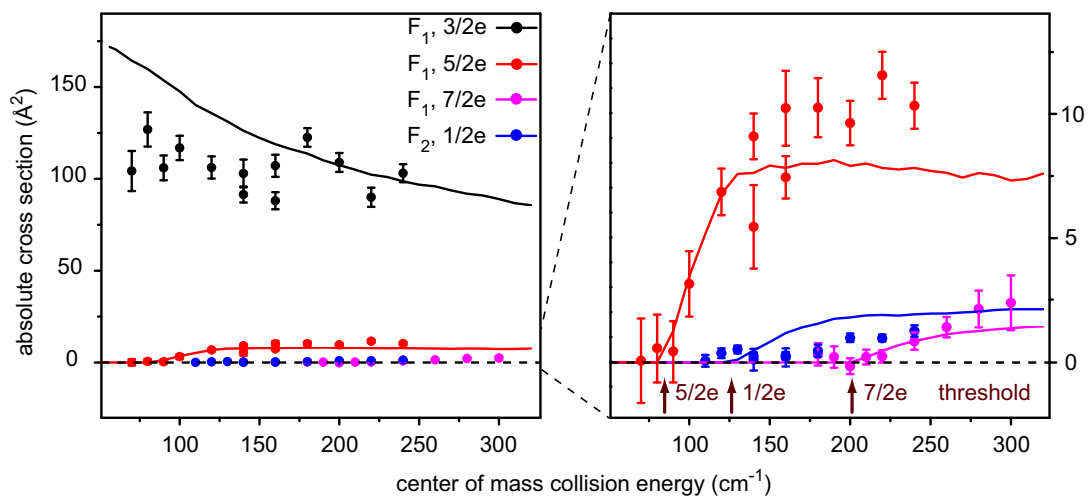}
\caption{Comparison of the collision energy dependence of the measured (data points with error bars) and calculated (solid curves) state-to-state inelastic scattering cross sections of OH $F_1(3/2f)$ radicals in collision with NO $F_1(1/2f)$ radicals. The cross sections were first measured relatively with respect to each other, and the vertical error bars indicate combined estimates of both statistical and systematic errors (2$\sigma$) \cite{Note:SI}. The vertical axis was then put on an absolute scale by a measurement of the absolute cross section for the $F_1(3/2e)$ channel at a collision energy of 220~cm$^{-1}$. The cross sections were computed on an energy grid of 10, 20, 30, 40,..., 320 cm$^{-1}$. The cross section for the dominant $F_1(3/2e)$ channel converged to within a few percent; the cross sections for the weaker channels vary by 20 to 50 \% depending on changes in the short-range part of the theoretical model \cite{Note:SI}.}
\end{figure}

\end{document}